\documentclass[sigconf]{acmart}

\AtBeginDocument{%
  \providecommand\BibTeX{{%
    \normalfont B\kern-0.5em{\scshape i\kern-0.25em b}\kern-0.8em\TeX}}}

\setcopyright{acmcopyright}
\acmDOI{10.1145/1122445.1122456}

\begin{document}

%%
%% The "title" command has an optional parameter,
%% allowing the author to define a "short title" to be used in page headers.
\title{Leaf-FM: A Learnable Feature Generation Factorization Machine for Click-Through Rate Prediction}

%%
%% The "author" command and its associated commands are used to define
%% the authors and their affiliations.
%% Of note is the shared affiliation of the first two authors, and the
%% "authornote" and "authornotemark" commands
%% used to denote shared contribution to the research.

\author{Qingyun She, Zhiqiang Wang, Junlin Zhang}
\affiliation{
     \institution{Sina Weibo Corp}
     \city{Beijing}
     \country{China}}
\email{qingyun_she@163.com, roky2813@sina.com, junlin6@staff.weibo.com}

%%
%% By default, the full list of authors will be used in the page
%% headers. Often, this list is too long, and will overlap
%% other information printed in the page headers. This command allows
%% the author to define a more concise list
%% of authors' names for this purpose.
%%\renewcommand{\shortauthors}{Trovato and Tobin, et al.}

%%
%% The abstract is a short summary of the work to be presented in the
%% article.
\begin{abstract}
Click-through rate (CTR) prediction plays important role in personalized advertising and recommender systems. Though many models have been proposed such as FM, FFM and DeepFM in recent years, feature engineering is still a very important way to improve the model performance in many applications because using raw features can rarely lead to optimal results. For example, the continuous features are usually transformed to  the  power forms by adding a new feature to allow it to easily form non-linear functions of the feature. However, this kind of feature engineering heavily relies on peoples experience and it is both time consuming and labor consuming. On the other side, concise CTR model with both fast online serving speed and good model performance is critical for many real life applications. In this paper, we propose LeafFM model based on FM to generate new features from the original feature embedding by learning the transformation functions automatically. We also design three concrete Leaf-FM models according to the different strategies of combing the original and the generated features. Extensive experiments are conducted on three real-world datasets and the results show Leaf-FM model outperforms standard FMs by a large margin. Compared with FFMs, Leaf-FM can achieve significantly better performance with much less parameters. In Avazu and Malware dataset, add version Leaf-FM  achieves comparable  performance with some deep learning based models such as DNN and AutoInt. As an improved FM model, Leaf-FM has the same computation complexity with FM in online serving phase and it means Leaf-FM is applicable in many industry applications because of its better performance and high computation efficiency.
\end{abstract}

%%
%% The code below is generated by the tool at http://dl.acm.org/ccs.cfm.
%% Please copy and paste the code instead of the example below.
%%


%%
%% Keywords. The author(s) should pick words that accurately describe
%% the work being presented. Separate the keywords with commas.

%%
%% This command processes the author and affiliation and title
%% information and builds the first part of the formatted document.
\maketitle

\section{Introduction}
Click-through rate (CTR) prediction is to predict the probability of a user clicking on the recommended items. It plays important role in personalized advertising and recommender systems. A series of works have been proposed to resolve this problem such as Logistic Regression (LR) \cite{10.1145/2487575.2488200}, Polynomial-2 (Poly2) \cite{rendle2010factorization}, tree-based models \cite{he2014practical}, tensor-based models \cite{koren2009matrix}, Bayesian models \cite{graepel2010web}, and Field-aware Factorization Machines (FFMs) \cite{juan2016field}. In recent years, Deep neural networks have been used as a powerful machine learning tool and achieve great success in this field for its strong expressive ability\cite{zhang2016deep,cheng2016wide,xiao2017attentional,guo2017deepfm,lian2018xdeepfm,qu2016product,wang2017deep}.

However, feature engineering is still a very important way to improve the model performance in these systems because using raw features can rarely lead to optimal results. In order to generate better performance models, usually a lot of work on the transformation of raw features is needed. For example, proper normalization of continuous features was critical for model convergence and the continuous features are usually transformed to the power forms by adding a new feature to allow it to easily form non-linear or sub-linear functions of the feature\cite{covington2016deep}. Say, in addition to the raw normalized feature $x$, we also input powers $x^2$ and $\sqrt{x}$ as new features into model. However, this kind of feature engineering heavily relies on people’s experience and it is both time consuming and labor consuming.

Besides the continuous feature, categorical feature is also commonly used in most CTR tasks, most of which are highly sparse. It's hard to transform these categorical feature to another new one by hand as continuous feature does. In order to promote the model's expressive ability and reduce the engineering resources, it's  necessary to make these feature generation  work to be automatic .

On the other side, production ranking systems require processing requests with high throughput under strict latency constraints. Large amount of machine resource and extra works are needed  even for simple DNN CTR model used by Facebook \cite{naumov2019deep, gupta2020architectural} in order to meet these application requirements. So simple models such as LR\cite{10.1145/2487575.2488200} or FM\cite{rendle2010factorization} show advantage over complex DNN ranking models under this consideration and are preferred in many real life applications. However,  DNN models usually achieve better performance than simple models. So concise CTR model which both has fast online serving speed and good model performance is critical for many real life applications.

In this work, we propose Leaf-FM model based on FMs to generate new features from the original feature embedding by learning the transformation functions automatically, which greatly enhances the expressive ability of model and saves the cost of feature engineering. Extensive experiments are conducted on three real-world datasets and the results show Leaf-FM model outperforms standard FMs by a large margin. In two of the three datasets we used, Leaf-FM  can achieve comparable  performance with  DNN models. As an improved FM model, Leaf-FM has the same computation complexity with FM in online serving phase and much better performance than FM, which means Leaf-FM is applicable in many  industry applications because of its better performance and high computation efficiency.

The contributions of our work are summarized as follows:

\begin{enumerate}
    \item We propose a novel Feature Generation Network (FGNet) to learn to automatically generate new features from the original one during model training, including both the continuous feature and categorical features. FGNet can greatly reduce the need for feature engineering in many real-world CTR tasks.

    \item We propose a new model named LEArnable Feature-generation Factorization Machine(Leaf-FM) which uses FGNet to boost the FM model's performance. We design three specific Leaf-FM models according to the different strategies of combing the original and the generated features: add version(LA-FM), sum version(LS-FM) and product version(LP-FM). The experiments show that the automatically generated features indeed boost the FM model’s performance significantly.

    \item We conduct extensive experiments on three real-world datasets and the experiment results show that Leaf-FM model outperforms standard FMs by a large margin. Compared with FFMs, Leaf-FM can achieve significantly better performance with much less parameters. On Avazu and Malware datasets, Leaf-FM can achieve comparable  performance with the DNN and DeepFM models.
\end{enumerate}

The rest of this paper is organized as follows. Section 2 introduces some related works which are relevant with our proposed model. Some preliminaries are described in Section 3 for easy understanding of the proposed model. We introduce our proposed Leaf-FM model in detail in Section 4. The experimental results on Criteo and Avazu datasets are presented and discussed in Section 5. Section 6 concludes our work in this paper.

\section{Related Work}

Factorization Machines (FMs) \cite{rendle2010factorization} and Field-aware Factorization Machines (FFMs) \cite{juan2016field} are two of the most successful CTR models. Other linear FM-based models are proposed, such as CoFM \cite{hong2013co}, FwFM \cite{pan2018field} and importance-aware FM \cite{oentaryo2014predicting}. However, these models show limited effectiveness in mining high-order latent patterns or learning quality feature representations.

Another line of research on FM-based models is to incorporate deep neural networks (DNNs). For example, Factorization-Machine Supported Neural Networks (FNN)\cite{zhang2016deep}, as well as the product-based neural network (PNN)\cite{qu2016product}, are feed-forward neural networks using FM to pre-train the embedding layer. More recently, hybrid architectures are introduced in Wide\&Deep \cite{cheng2016wide}, DeepFM \cite{guo2017deepfm} and xDeepFM \cite{lian2018xdeepfm} by combining shallow components with deep ones to capture both low- and high-order feature interactions. Wide \& Deep Learning\cite{xiao2017attentional} jointly trains wide linear models and deep neural networks to combine the benefits of memorization and generalization for recommender systems. DeepFM \cite{guo2017deepfm} replaces the wide part of Wide \& Deep model with FM and shares the feature embedding between the FM and deep component. FiBiNET\cite{huang2019fibinet} can dynamically learn feature importance via the Squeeze-Excitation network (SENET) mechanism and feature interactions via bilinear function. The eXtreme Deep Factorization Machine (xDeepFM) \cite{lian2018xdeepfm} also models the low-order and high-order feature interactions in an explicit way by proposing a novel Compressed Interaction Network (CIN) part. AutoInt \cite{song2019autoint} uses a multi-head self-attentive neural network to explicitly model the feature interactions in the low-dimensional space. However, DNN models significantly increase the computation time and for some complex models it’s hard to deploy them in real-world applications.

\section{Preliminaries}

\subsection{Factorization Machines and Field-aware Factorization Machine}
Factorization Machines (FMs) \cite{rendle2010factorization} is one of the most widely used CTR models in many real-world applications because of its conciseness.

As we all know, FMs model interactions between features $i$ and $j$ as the dot products of their corresponding embedding vectors as follows:

\begin{equation}
  \hat{y}(x) = w_0 + \sum^m_{i=1}w_ix_i + \sum^{m}_{i=1}\sum^{m}_{j=i+1}\left<v_i,v_j\right>x_ix_j
\end{equation}

An embedding vector $v_i \in \mathbb{R}^d$ for each feature is learned by FM, $d$ is a hyper-parameter which is usually a small integer and $m$ is the feature number. However, FM neglects the fact that a feature might behave differently when it interacts with features from other fields. To explicitly take this difference into consideration, Field-aware Factorization Machines (FFMs) learn extra $f-1$ embedding vectors for each feature(here $f$ denotes field number):

\begin{equation}
  \hat{y}(x) = w_0  + \sum^m_{i=1}w_ix_i + \sum^{m}_{i=1}\sum^{m}_{j=i+1}\left<v_{ij},v_{ji}\right>x_ix_j
\end{equation}

where $v_{ij} \in \mathbb{R}^d$ denotes the embedding vector of the $j$-th entry of feature $i$ when feature $i$ is interacting with fields $j$. $d$ is the
embedding size.

\subsection{Layer Normalization}
Normalization techniques have been recognized as very effective components in deep learning. For example, Batch Normalization (Batch Norm or BN)\cite{ioffe2015batch} normalizes the features by the mean and variance computed within a mini-batch. This has been shown by many practices to ease optimization and enable very deep networks to converge. Another example is layer normalization (Layer Norm or LN)\cite{ba2016layer} which was proposed to ease optimization of recurrent neural networks. Statistics of layer normalization are not computed across the $N$ samples in a mini-batch but are estimated in a layer-wise manner for each sample independently. Specifically, let $x = (x_1, x_2, ..., x_H)$ denotes the vector representation of an input of size $H$ to normalization layers. LayerNorm re-centers and re-scales input $\mathbf{x}$ as

\begin{equation}
  \setlength\abovedisplayskip{0pt}
  \begin{split}
  &\mathbf{h} = \mathbf{g} \odot N(\mathbf{x}) + \mathbf{b}, \quad N(\mathbf{x}) = \frac{\mathbf{x}-\mu}{\delta}, \\
  & \mu = \frac{1}{H}\sum^H_{i=1}x_i, \quad \delta = \sqrt{\frac{1}{H}\sum^{H}_{i=1}(x_i - \mu)^2}
\end{split}
\end{equation}
where $h$ is the output of a LayerNorm layer. $\odot$ is an element-wise production operation. $\mu$ and $\delta$ are the mean and standard deviation of input. Bias $\mathbf{b}$ and gain $\mathbf{g}$ are parameters with the same dimension $H$.

\section{Our Proposed Model}
As mentioned in the section 1, feature engineering such as transforming the input continuous features to the sub-linear or super-linear form helps promote the model performance. However, the experience about how to transform the continuous feature needs expert knowledge. In real-world applications, considerable user's demographics and item's attributes are usually categorical. It seems hard to adopt the similar feature transformation on categorical feature.

In this paper, we intent to automatically learn the transformed new feature from the original one and boost FM model's expressive ability by combing those new generated feature. We learn the generated new feature through a network on the feature embedding.

As all we know, features in CTR tasks usually can be segregated into the following two groups.

\begin{enumerate}
  \item categorical features. This type of feature is common and the one-hot representation may produce very sparse features. We usually map one-hot representation to dense, low-dimensional embedding vectors suitable for complex transformation and these embedding carry richer information than one-hot representations. We can obtain feature embedding $v_i$ for one-hot vector $x_i$ via:
  \begin{equation}
    v_i = W_ex_i
  \end{equation}
where $W_e \in\mathbb{R}^{d\times n}$ is the embedding matrix of $n$ features and $d$ is the dimension of field embedding.

  \item Numerical features. There are two widely used approaches to convert the numerical feature into embedding. The first one is to quantize each numerical feature into discrete buckets, and the feature is then represented by the bucket ID. We can map  bucket ID to an embedding vector. The second method maps the feature field into an embedding vector as follows:

  \begin{equation}
    v_i = e_ix_i
  \end{equation}

where $e_i$ is an embedding vector for field $i$ with size $d$, and $x_i$ is a scalar value which means the actual value of that numerical feature. In our experiments, we adopt the second approach to convert numerical features into embedding.
\end{enumerate}

\begin{figure}[!]
  \centering
  \includegraphics[width=0.5\linewidth]{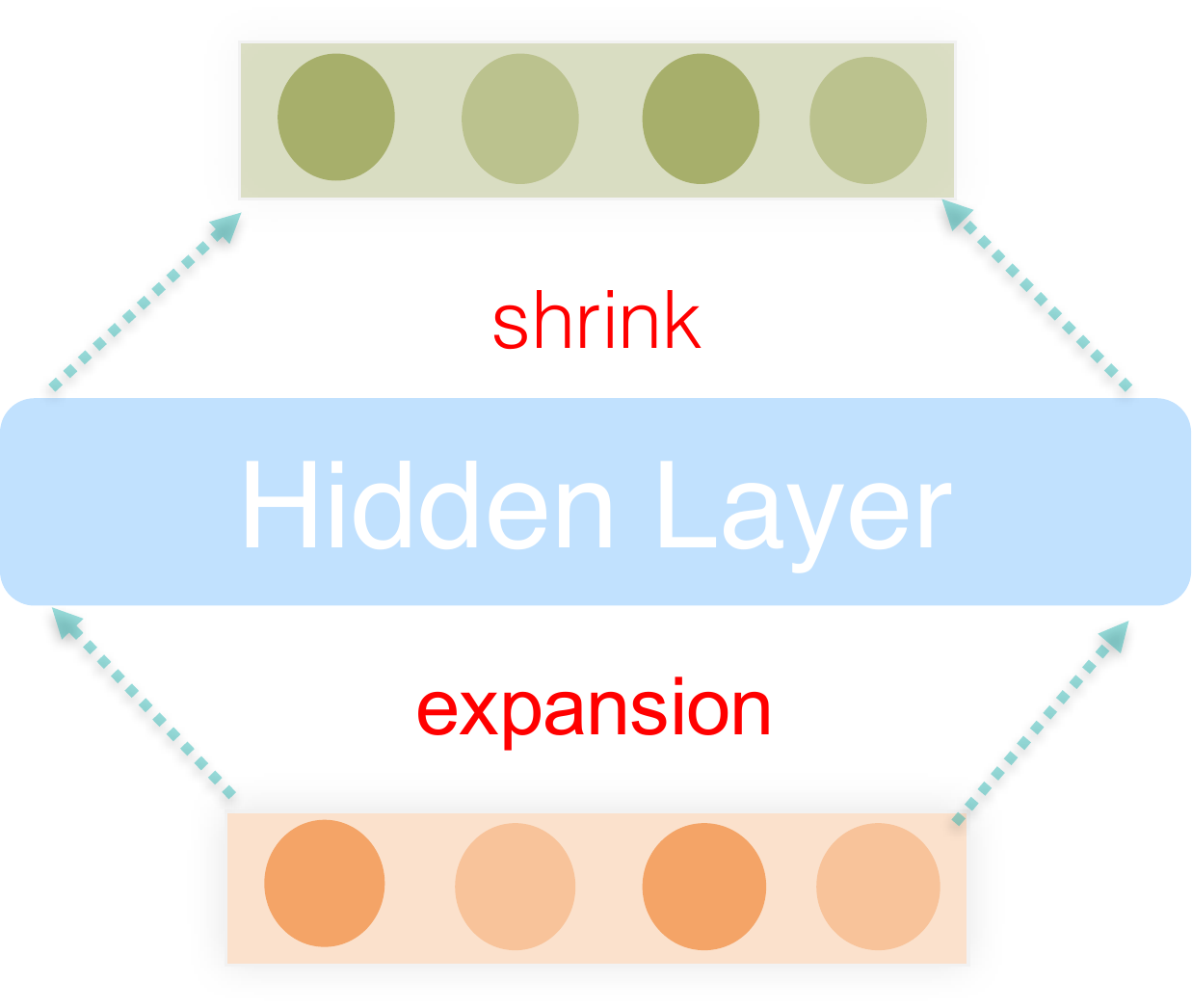}
  \caption{Structure of Feature Generation Network}
  \label{Fig.fgnet}
\end{figure}

\begin{figure*}[!h]
  \centering
  \includegraphics[width=0.79\linewidth]{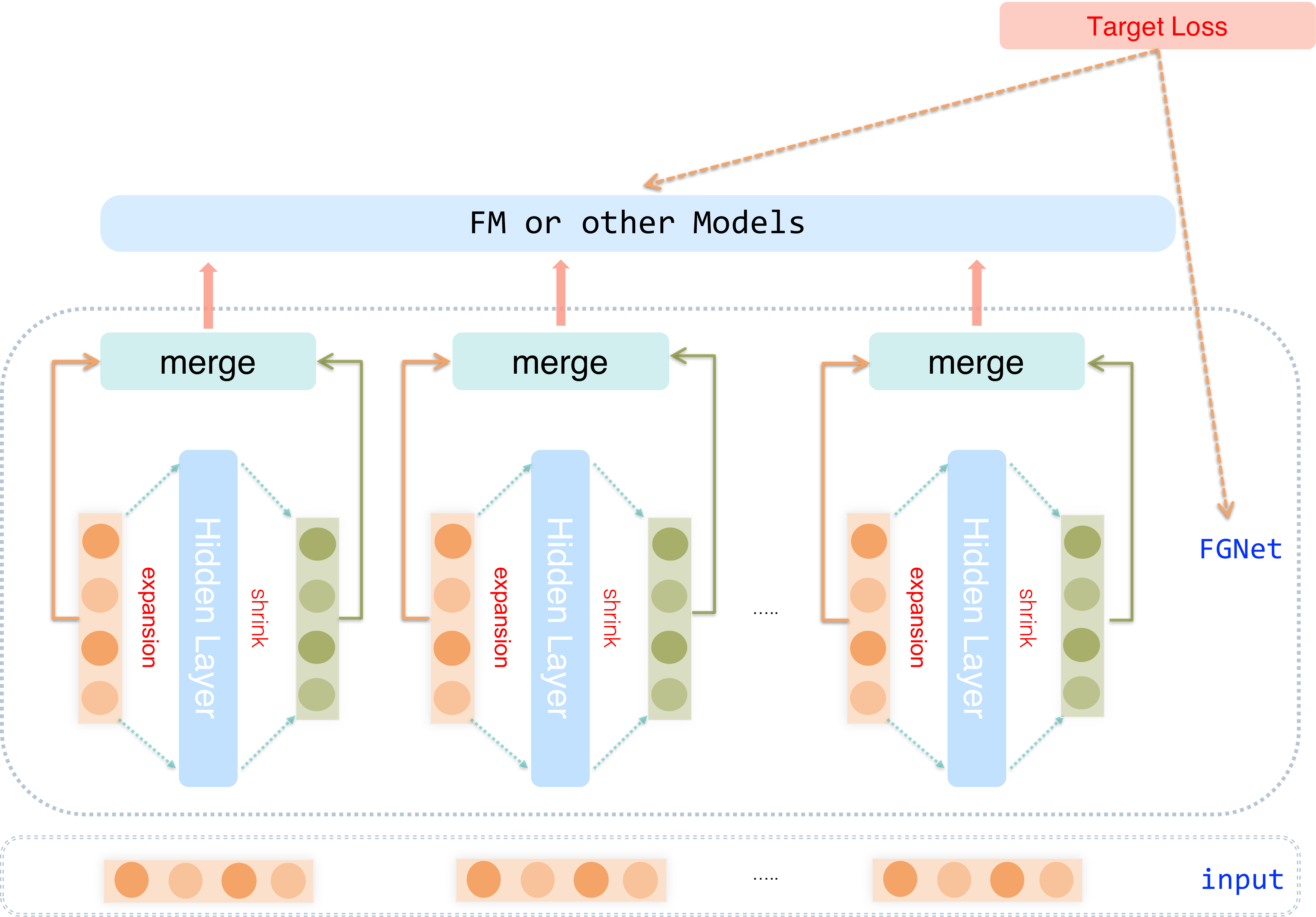}
  \caption{Structure of Leaf-FM}
  \label{Fig.leaffm}
\end{figure*}
%[width=0.79\linewidth,height=8.5cm]
\subsection{Learn to Generate New Feature}

In order to enhance the model’s expressive ability by learning to generate new features from the original one, we propose a Feature Generation Network(FGNet) in this work as depicted in Figure \ref{Fig.fgnet}.

Suppose a feature embedding $v_i$ is one feature in field $k$ . To generate the new feature from $v_i$, two fully connected (FC) layers are used on feature $v_i$. The first FC layer is a dimensionality-expansion layer with parameters $W_{k1}$ and expansion ratio $r$ means the size of FC layer is $rd$ if the vector size is $d$. Expansion ratio is a hyper-parameter and we will discuss its effect in detail in Section 5.4. The second FC layer decreases dimensionality with parameters $W_{k2}$, which is equal to dimension of feature embedding $v_i$. Formally, the feature generation process is calculated as follows:

\begin{equation}
g_i = \mathcal{F}_{FGNet}(v_i, W_k) = \delta(W_{k2} \cdot \delta(W_{k1}\cdot v_i))
\end{equation}

where $\delta$ refers to the non-linear function, $W_{k1} \in \mathbb{R}^{rd\times d}$ and
$W_{k2} \in \mathbb{R}^{d\times rd}$ are parameters for two FC layers. The size of generated feature $g_i$ is $d$, which is equal to the size of original embedding $v_i$.

Notice here that each feature field maintains its parameters $W_k$ and the total number of parameters for feature generation is $f \ast W_k$. $f$ is the number of fields. New FGNets are used if we want to produce more than one generated feature for $v_i$. It's not hard to find that FGNet is a rather small network with little parameters because we usually adopt small feature embedding size in real life applications.

Though the above-mentioned feature generation module has only two FC layers, we can stack FC layers to add further non-linear transformation for each original feature vector $v_i$ in order to generate better new features. So the depth of feature generation module can be regarded as a hyper-parameter. We find that the performance of Leaf-FM increases with deeper depth in feature generation module and we will discuss this in detail in Section 5.4.

In this way, this network can be trained to learn to transform original feature to a super-linear or sub-linear form, which can be added into models as new features. Notice that it’s a general approach to learn to generate new feature and doesn’t relies on concrete model we used.

\subsection{Leaf-FM}
In this paper, we propose Leaf-FM model as depicted in Figure \ref{Fig.leaffm} to show how to combine this feature generation module and FM model to improve the baseline model's performance. As mentioned above, we can learn to generate one or more new transformed features $g_i$ for each feature $v_i$ and there are various merging strategies to combine the original feature and generated features during model training in the FM models. We propose three version of Leaf-FM according to the different feature merging methods in this work.

Suppose we use the FGNet generate some new features for each feature $v_i$, we have the following feature set for $v_i$:

\begin{equation}
FS_{v_i} = \left\{v_i, g_{i1}, ..., g_{ij}, ..., g_{iu}\right\}
\end{equation}

where $g_{ij}$ is the $j$-th generated feature from feature $v_i$, $u$ is the number of new generated features for each feature $v_i$.

\subsubsection{Add Version Leaf-FM}
A direct method to merge the newly generated features in FM is to regard these features as independent new features and directly add them into the model. So the feature numbers of FM increase from m to $(u+1)m$ if we generate $u$ new feature $g_i$ for each feature $v_i$. We call this version Leaf-FM 'Add version Leaf-FM'(LA-FM) which has the following format:

\begin{equation}
      \hat{y}(x) = w_0  + \sum^m_{i=1}w_ix_i + \sum^{(u+1)m}_{i=1}\sum^{(u+1)m}_{j=i+1}\left<v_i,v_j\right>x_ix_j
\end{equation}

For this version Leaf-FM, we use either Relu or identity as non-linear function($\Phi(x)$) in feature generation network:

%\begin{equation}
%  g_i = \mathcal{F}_{FGNet}(v_i, W_k) = Relu(W_{k2}Relu(W_{k1}v_i + %\mathbf{\beta}_{k1}) + \mathbf{\beta}_{k2})
%\end{equation}

\begin{equation}
  g_i = \mathcal{F}_{FGNet}(v_i, W_k, \beta_k) = \Phi(W_{k2}\Phi(W_{k1}v_i + \mathbf{\beta}_{k1}) + \mathbf{\beta}_{k2})
\end{equation}

\subsubsection{Sum Version Leaf-FM}
Another merging approach is to sum all the features in feature set $FS_{vi}$ to one vector as follows:
\begin{equation}
  \mathcal{F}_s(v_i) = v_i + \sum_{j=1}^u g_{ij}
\end{equation}

where $\mathcal{F}_s(v_i)$ is the merged new feature vector which is equal to the size of feature $v_i$, $g_{ij}$ is the $j$-th generated feature from feature $v_i$, $u$ is the number of new generated features for each feature $v_i$. We call Leaf-FM with this kind of merging strategy 'Sum Version Leaf-FM'(LS-FM) and LS-FM model has the following format:

\begin{equation}
  \hat{y}(x) = w_0  + \sum^m_{i=1}w_ix_i + \sum^{m}_{i=1}\sum^{m}_{j=i+1}\left<\mathcal{F}_{s}(v_i),\mathcal{F}_s(v_j)\right>x_ix_j
\end{equation}

For this version Leaf-FM, we also use Relu as non-linear function in feature generation network:

\begin{equation}
  g_i = \mathcal{F}_{FGNet}(v_i, W_k, \beta_k) = Relu(W_{k2}Relu(W_{k1}v_i + \beta_{k1}) + \beta_{k2})
\end{equation}

\begin{table*}
%\centering
\caption{A summary of model complexities (ignoring the bias term) and computation efficiency. m and f are feature number and field number respectively, d is the embedding vector dimension, and r is the expansion ratio of FC for Leaf-FM models.}
%\resizebox{\linewidth}{!}{
\begin{tabular}{lccc}
\toprule
 Models   & \textbf{Number of parameters} &  \textbf{Offline Training} &  \textbf{Online Serving} \\
  \midrule
  FM       & $m + m\ast d$  & $O(m\ast d)$ & $O(m\ast d)$     \\
  FFM       & $m + m \ast (f-1)\ast d$  & $O(m^2\ast d)$ & $O(m^2\ast d)$     \\
  Leaf-FM & $m + m\ast d + f\ast 2(rd\ast d)$ & $O(f\ast r\ast d^3 + m\ast d)$ & $O(m\ast d)$ \\
\bottomrule
\end{tabular}
\label{tab:modelcomp}
\end{table*}

\subsubsection{Product Version Leaf-FM}
In order to increase the non-linearity of the merging process, we can merge the new generated feature $g_i$ and original feature $v_i$ by element-product and normalization as follows:

\begin{equation}
  \mathcal{F}_p(v_i) = LayerNorm(g_i \odot v_i)
\end{equation}

where $g_i$ is the generated feature from feature $v_i$, $\odot$ is the element-wise product of two vectors, the layer normalization is used on the results of element-wise product to further increase the non-linear ability of the generating model.

We call Leaf-FM with this merging strategy 'Product Version Leaf-FM'(LP-FM) and LP-FM model has the following format:

\begin{equation}
    \hat{y}(x) = w_0  + \sum^m_{i=1}w_ix_i + \sum^{m}_{i=1}\sum^{m}_{j=i+1}\left<\mathcal{F}_{p}(v_i),\mathcal{F}_p(v_j)\right>x_ix_j
\end{equation}

For this version Leaf-FM, we use identity function in feature generation network:

\begin{equation}
  g_i = \mathcal{F}_{\tiny{FGNet}}(v_i, W_k, \beta_k) = W_{k2}(W_{k1}v_i + \beta_{k1}) + \beta_{k2}
\end{equation}

Notice that each feature $v_i$ can  produces only one new feature $g_i$ in this version of Leaf-FM because of the dot product operation. While the number of the generated new features for each feature $v_i$ in add-version and sum-version Leaf-FM can be a hyper-parameter and we will discuss this in detail in Section 5.2.

\subsection{Joint Training of Feature Generation Module and FM}
Since the Leaf-FM has two parts of parameters to train: feature vectors in FM and the FC network parameters in feature generation module, so we can train them jointly as shown in Figure \ref{Fig.leaffm}.

For binary classifications, the log loss  is used as loss function in our work and the optimization process is to minimize the following objective function:

%\pounds
\begin{equation}
  \mathcal{L} = -\frac{1}{N}\sum^N_{i=1}y_i\log(\hat{y}_i)+(1-y_i)\log(1-\hat{y}_i) + \lambda \|\Theta\|
\end{equation}

where $N$ is the total number of training instances, $y_i$ is the ground truth of $i$-th instance and $\hat{y}_i$ is the predicted CTR.  $\lambda$ denotes the regularization term and $\Theta$ denotes the set of parameters, including these in feature vector matrix and feature generation module.

%\begin{table}
%\centering
%\caption{A summary of model complexities (ignoring the bias term) and computation %efficiency. m and f are feature number and field number respectively, d is the %embedding vector dimension, and r is the reduction ratio of FC for Leaf-FM models.}
%\resizebox{\linewidth}{!}{
%\begin{tabular}{lccc}
%\hline
%         & Number of &  Offline &  Online \\
%Models  &  parameters &  Training &  Serving \\
%\hline
%FM       & m + m\ast d  & O(m\ast d) & O(m\ast d)     \\
%FFM       & m + m \ast (f-1)\ast d  & O(m\ast 2d) & O(m\ast 2d)     \\
%Leaf-FM & m + m\ast k + f\ast 2(rd\ast d) & O(m\ast 2d) & O(m\ast d) \\
%\hline
%\end{tabular}}
%\label{tab:complexities}
%\end{table}

\subsection{Model Complexity and Computation Efficiency}
FM is widely used in many industry applications because of its high computation efficiency. As shown in Table \ref{tab:modelcomp}, the number of parameters in FM is $m + m\ast d$, where $m$ accounts for the weights for each feature in the linear part $\{w_i |i = 1, ...,m\}$ and $m\ast d$ accounts for the embedding vectors for all the features $\{v_i |i = 1, ...,m\}$. FM has a $O(m\ast d)$ linear runtime, which makes it applicable from a computational point of view.

In recent years, many new models such as FFM and DeepFM have been proposed. Though the performances are boosted greatly , many new models's high model complexity hinders their real-word application. For example, the number of parameters of FFM is $m + m\ast (f$ - 1$)\ast d$ since each feature has $f$ - $1$ embedding vectors and it has a $O(m^2\ast d)$ computation complexity.

Leaf-FM use $f\ast 2(rd\ast d)$ additional parameters for generating new features so that the total number of parameters of Leaf-FM is $m + m \ast d + f \ast 2(rd \ast d)$. The extra parameter number introduced by Leaf-FM is very small because we usually use small feature embedding size.  The parameter number of Leaf-FM is comparable with that of FM and significantly less than that of FFM. Though the offline training of Leaf-FM needs $O(f\ast r\ast d^3 + m\ast d)$ runtime in order to jointly train the feature generation module, Leaf-FM has the same computation complexity with FM in online serving phase. After training the Leaf-FM, we can transform the original feature $v_i$ to $F_s(v_i)$ for each feature in offline phase. So the standard FM can be used in online serving phase and that means our proposed Leaf-FM model is applicable in industry applications because of its better model’s performance and high computation efficiency.

\section{Experimental Results}
To comprehensively evaluate our proposed method, we design some experiments to answer the following research questions:

\begin{itemize}

\item\noindent\textbf{RQ1} Are the new features generated by Feature Generation Network useful for FM model? How does the number of the newly generated features influence the model performance?

\item\noindent\textbf{RQ2} Can our proposed several Leaf-FM models outperform complex models like FFM? How about the performance comparison even with the deep learning based models such as DNN and DeepFM?

\item\noindent\textbf{RQ3} How does the hyper-parameters of Feature Generation Network influence the model performance?

\item\noindent\textbf{RQ4} How does the hyper-parameters of Leaf-FM influence the model performance?

\end{itemize}

In the following, we will first describe the experimental settings, followed by answering the above research questions.

\subsection{Experiment Setup}

%\subsubsection{Datasets}
\paragraph{Datasets}

The following three data sets are used in our experiments:

\begin{enumerate}
  \item \textbf{Criteo\footnote{Criteo \url{http://labs.criteo.com/downloads/download-terabyte-click-logs/}} Dataset:}
  As a very famous public real world display ad dataset with each ad display information and corresponding user click feedback, Criteo data set is widely used in many CTR model evaluation. There are 26 anonymous categorical fields and 13 continuous feature fields in Criteo data set.

  \item \textbf{Avazu\footnote{Avazu \url{http://www.kaggle.com/c/avazu-ctr-prediction}} Dataset:}
    The Avazu dataset consists of several days of ad click- through data which is ordered chronologically. For each click data, there are 24 fields which indicate elements of a single ad impression.

  \item \textbf{Malware\footnote{Malware \url{https://www.kaggle.com/c/microsoft-malware-prediction}} Dataset:}
  Malware is a dataset from Kaggle competitions published in the Microsoft Malware prediction. The goal of this competition is to predict a Windows machine’s probability of getting infected by various families of malware, based on different properties of that machine. The malware prediction task can be formulated as a binary classification problem like a typical CTR estimation task does.
\end{enumerate}

%\paragraph{Criteo\footnote{Criteo \url{http://labs.criteo.com/downloads/download-terabyte-click-logs/}} Dataset}

%  As a very famous public real world display ad dataset with each ad display information and corresponding user click feedback, Criteo data set is widely used in many CTR model evaluation. There are 26 anonymous categorical fields and 13 continuous feature fields in Criteo data set.

%\paragraph{Avazu\footnote{Avazu http://www.kaggle.com/c/avazu-ctr-prediction} Dataset}

%  The Avazu dataset consists of several days of ad click- through data which is ordered chronologically. For each click data, there are 24 fields which indicate elements of a single ad impression.

%\paragraph{Malware \footnote{Malware https://www.kaggle.com/c/malware-classification} Dataset}

%Malware is a dataset from Kaggle competitions published in the Microsoft Malware Classification Challenge. The goal of this competition is to predict a Windows machine’s probability of getting infected by various families of malware. It is almost half a terabyte when uncompressed and consists of disassembly and bytecode malware files representing a mix of 9 different families.

We randomly split instances by 8:1:1 for training , validation and test while Table \ref{tab:datasets} lists the statistics of the evaluation datasets. For these datasets, a small improvement in prediction accuracy is regarded as practically significant because it will bring a large increase in a company’s revenue if the company has a very large user base.

\begin{table}[h]
\centering
\caption{Statistics of the evaluation datasets}
\begin{tabular}{lccc}
\toprule
Datasets  & \#Instances & \#fields & \#features \\
\midrule
Criteo       & 45M  & 39 & 30M     \\
Avazu       & 40.43M  & 24 & 9.5M     \\
Malware     & 8.92M  & 82 & 0.97M \\
\bottomrule
\end{tabular}
\label{tab:datasets}
\end{table}

%\subsubsection{Evaluation Metrics}
\paragraph{Evaluation Metrics}

AUC (Area Under ROC) is used in our experiments as the evaluation metrics. This metric is very popular for binary classification tasks. AUC’s upper bound is $1$ and larger value indicates a better performance.

%\subsubsection{ModelsforComparisons}
\paragraph{Models for Comparisons}

We compare the performance of the FM, FFM, DNN, DeepFM and AutoInt models as baseline and  all of which are discussed in Section 2 and Section 3.

%\subsubsection{Implementation Details}
\paragraph{Implementation Details}

We implement all the models with Tensorflow in our experiments. For optimization method, we use the Adam with a mini-batch size of $1024$ and a learning rate is set to $0.0001$. We make the dimension of field embedding for all models to be a fixed value of $10$ for FM-related models and $8$ for FFM model which was adopted in \cite{juan2016field}. Notice that large embedding size for FFM leads to over-fitting because the FFM model has too much parameters. For models with DNN part, the depth of hidden layers is set to $3$, the number of neurons per layer is $400$, all activation function are ReLU. We conduct our experiments with $2$ Tesla $K40$ GPUs.

\subsection{Usefulness of the Generated Features(RQ1)}
The overall performance for CTR prediction of different models on three real-world datasets is shown in Table \ref{tab:overalperformance}. Compared with the performance of FM model, the three versions of Leaf-FM outperform FM by a large margin almost on all datasets.  Because  the  only difference between the FM and Leaf-FM model is that Leaf-FM uses  the new features generated by feature generation network, we can draw the conclusion that the generated features by FGNet can greatly increase the simple model’s expressive ability by automatically learning to add some transformed sub-linear and super-linear new features. The only exception is the performance of product version Leaf-FM model on Malware dataset and this indicates how to merge the original features and generated features still matters on different dataset.

\begin{table}
\centering
\caption{Overall performance (AUC) of different models on three datasets(d=4 for FFM and 10 for all other model, $d$ means embedding size, $l$ means depth of DNN, $n$ means the size of FC layer of DNN, $p$ means network depth of FGNet, $g$ means number of the generated feature; $e$ means expansion rate)}
%\resizebox{\linewidth}{!}{
\begin{tabular}{l|cccc}
\toprule
  & \textbf{Avazu}  & \textbf{Malware}  & \textbf{Criteo}& \\
\midrule
Model & AUC & AUC & AUC &Setting \\
FM       & 0.7757    & 0.7225  & 0.7896 & $d$=10    \\
FFM & 0.7791 & 0.7240 & 0.7951  & $d$=8 \\
\midrule
DNN  & 0.7820 & 0.7263 & 0.8054  & $l$=3;$n$=400;$d$=10 \\
DeepFM & 0.7833  & 0.7295 & 0.8056 & $l$=3;$n$=400;$d$=10 \\
AutoInt & 0.7824 & 0.7282 & 0.8051  & $b$=3;$d$=10 \\
\midrule
  LA-FM & 0.7822 & 0.7281 & 0.7993 & $e$=1;$p$=2;$g$=1;$d$=10  \\
  LS-FM & 0.7803 & 0.7260 & 0.8003 & $e$=1;$p$=2;$g$=3;$d$=10  \\
  LP-FM & 0.7803 & 0.7167 & 0.8005 & $e$=1;$p$=2;$g$=1;$d$=10 \\
\bottomrule
\end{tabular}%}
\label{tab:overalperformance}
\end{table}

\begin{figure}
  \setlength{\abovecaptionskip}{1pt}
  \includegraphics[width=0.85\linewidth]{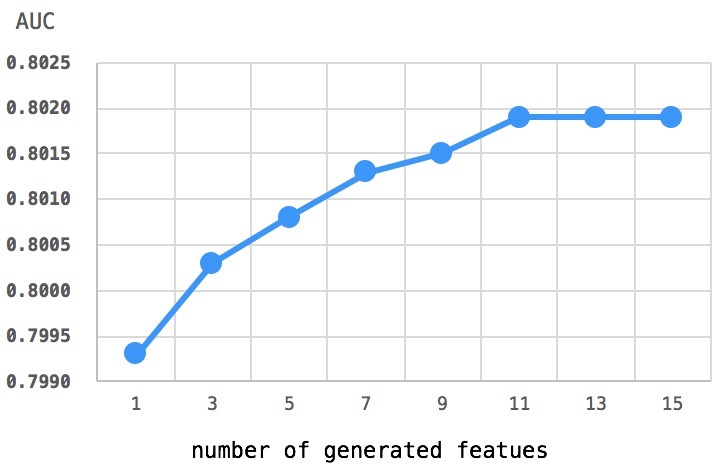}
  \caption{Performances of LA-FM with Different numbers of Generated Feature}
  \label{fig.fig3}
\end{figure}

We conduct some experiments to observe the influence of the number of newly generated features for each original feature on LA-FM model on Cretio dataset. The experimental results are shown in Figure \ref{fig.fig3}. From the results, we can see that the performance increases gradually with the increase of the number of newly generated features until the number reaches 11. This indicates that the more features generated by the FGNet, the more useful information we can get.

%[width=10pt, height=10pt]
%\begin{table}
%\centering
%\caption{Overall performance (AUC) of Add version LeafFM with different number of newly generated features for each original feature on Criteo  dataset(embedding size=10,network depth=2,expansion rate=1)}
%\resizebox{\linewidth}{!}{
%\begin{tabular}{l|cccccc}
%\toprule
% & \textbf{1} & \textbf{3} & \textbf{5} & \textbf{7} & \textbf{9} & \textbf{11} \\
%\midrule
%LA-FM &  0.7993 & 0.8003 & 0.8008 & 0.8013 & 0.8015 & 0.8019   \\
%\bottomrule
%\end{tabular}%}
%\label{tab:featureNum}
%\end{table}

\subsection{Performance Comparison(RQ2)}
From the experimental results shown in Table \ref{tab:overalperformance}, we have the following key observations:

\begin{enumerate}
  \item As an improved FM model, all three versions of Leaf-FM model achieve better performance on all three datasets and obtain significant improvements over baseline FM model. We also conduct a significance test to verify that our proposed models outperforms baselines with the significance level $\alpha$ = 0.01.
 \item Leaf-FM outperforms more complex models like FFM on almost all  three datasets. Considering Leaf-FM has much less parameters and nearly linear online runtime, we can see that Leaf-FM model is a much more applicable CTR model.
  \item As we all know, DNN models usually outperform FM with a large margin and our experimental results prove that. However, some version of Leaf-FM is comparable with DNN model on two datasets as shown in Table \ref{tab:overalperformance}. Compared with deep learning based CTR models such as DNN and AutoInt models, add version Leaf-FM (LA-FM) can achieve comparable performance on Malware and Avazu datasets. Though Leaf-FM underperforms DNN models on Criteo dataset, our experiments in Section 5.5 show the performance will greatly increase if we set embedding size larger for LA-FM.
  LA-FM even outperforms DNN model on Malware dataset and that's out of our expectation. Because DNN model use MLP to capture high-order feature interaction and FM-like model doesn't has this ability. This may suggests that some feature engineering work like feature transformation is important for some dataset  such as Malware and Avazu. The experimental results also prove that Leaf-FM model works well by automatically learning to generate new features and it can be used to reduce the human labor in feature engineering in many real-world CTR tasks.
  \item As for the comparison of three different versions of Leaf-FM, we can see from the results that feature merging approach still matters. Different tasks need specific feature merging strategy. For example, LS-FM and LP-FM outperform LA-FM on Cretio dataset while LA-FM and LS-FM are better choices for Malware dataset.
  \item As an improved FM model, Leaf-FM has the same computation complexity with FM in online serving phase. At the same time, Leaf-FM has comparable performance with some DNN models. These mean Leaf-FM is applicable in many  industry applications because of its better performance and high computation efficiency.
\end{enumerate}

\begin{table}
\centering
\caption{Overall performance (AUC) of Add version LeafFM with different expansion rate on Criteo dataset(embedding size=10,network depth=2,number of the generated feature=1)}
%\resizebox{\linewidth}{!}{
\begin{tabular}{l|cccccc}
\toprule
 & \textbf{1} & \textbf{2} & \textbf{3} & \textbf{4} & \textbf{5} & \textbf{7} \\
\midrule
LA-FM &  0.7993 & 0.7994 & 0.7996 & 0.8002 & 0.7980 & 0.7990   \\
\bottomrule
\end{tabular}%}
\label{tab:expansionrate}
\end{table}

\subsection{Hyper-Parameter  of FGNet(RQ3)}
In this section, we study the impact of hyper-parameters of Feature Generation Network on Leaf-FM model. For the FG network, expansion rate and network depth are two hyper-parameters to influence the model performance. The experiments are conducted on Criteo datasets via changing one hyper-parameter while holding the other settings.

\paragraph{Expansion Rate.}  We conduct some experiments to adjust the expansion rate in LA-FM model from $1$ to $7$,which means size of hidden layer is from $10$ to $70$ because the embedding size is set $10$. We list the results in Table \ref{tab:expansionrate}. We can observe a slightly performance increase with the increase of expansion rate until it’s greater than $4$.

\begin{table}[h]
\centering
\caption{Overall performance (AUC) of Add version LeafFM with different network depth on Criteo dataset(embedding size=10,expansion rate=1, number of the generated feature=1))}
%\resizebox{\linewidth}{!}{
\begin{tabular}{l|cccccc}
\toprule
 & \textbf{2} & \textbf{3} & \textbf{4} & \textbf{5} & \textbf{7} & \textbf{10} \\
\midrule
LA-FM &  0.7993 & 0.7991 & 0.7996 & 0.8003 & 0.8001 & 0.7946   \\
\bottomrule
\end{tabular}%}
\label{tab:networkdeep}
\end{table}

\paragraph{Depth of Network.} The results in Table \ref{tab:networkdeep} show the impact of number of hidden layers. We can observe that the performance of LA-FM increases with the depth of network at the beginning. However, model performance degrades when the depth of network is set greater than 5. Over-fitting of deep network is maybe the reason for this.

\subsection{Hyper-Parameter of Leaf-FM (RQ4)}
In this section, we study the impact of hyper-parameters on Leaf-FM model. For the FM or FFM models, the number of feature embedding size is our main concern. The experiments are conducted on Criteo datasets via changing embedding size from $10$ to $120$ while holding the other settings.

\paragraph{Embedding Size.} Table \ref{tab:embeddingsize} shows the experimental results . It can be seen that the number of feature embedding size greatly influences the model performance. With the increase of this number, the performance increases rapidly until the number reaches $100$. The AUC of the model with embedding size $100$ is $0.8055$,which outperforms the DNN model with the smaller embedding size.

\begin{table}
\centering
\caption{Overall performance (AUC) of Add version LeafFM with different embedding size on Criteo dataset(expansion rate=1,network depth=2,number of the generated feature=1)}
%\resizebox{\linewidth}{!}{
\begin{tabular}{l|cccccc}
\toprule
 & \textbf{10} & \textbf{30} & \textbf{50} & \textbf{80} & \textbf{100} & \textbf{120} \\
\midrule
LA-FM &  0.7993 & 0.8030 & 0.8048 & 0.8054 & 0.8055 & 0.8051   \\
\bottomrule
\end{tabular}%}
\label{tab:embeddingsize}
\end{table}

\section{Conclusion}

In this paper, we propose Leaf-FM model to generate new features from the original feature embedding by learning the transformation functions automatically, which greatly enhances the expressive ability of model and saves the cost of feature engineering.  Extensive experiments are conducted on real-world datasets and the results show Leaf- FM model outperforms standard FMs by a large margin. Leaf-FM even can achieve comparable performance with the DNN CTR models. As an improved FM model, Leaf-FM has the same computation complexity with FM in online serving phase and it means Leaf-FM is applicable in many industry applications because of its better model’s performance and high computation efficiency.

\bibliographystyle{ACM-Reference-Format}
\bibliography{ijcai20}

\end{document}